\documentclass[aps,amsmath,reprint,amssymb]{revtex4-1}
\usepackage{graphicx}
\usepackage[utf8]{inputenc}
\usepackage[T1]{fontenc}
\usepackage{lmodern}
\usepackage{dcolumn}
\usepackage{bm}
\usepackage{amsmath}

\begin{document}
\title{
Rapidity-dependent eccentricity scaling in relativistic heavy-ion collisions
}
\author{Rodrigo Franco and Matthew Luzum} 
\affiliation{
    Instituto de F\'{i}sica, Universidade de S\~{a}o Paulo, 
    R. do Mat\~{a}o 1371, 05508-090 S\~{a}o Paulo, SP, Brazil
}

	\begin{abstract}

	There is a well-established relation between the spatial asymmetry in the initial stage of a heavy-ion collision and the final momentum anisotropy, which allows for a separation of effects from initial conditions vs.~later evolution and has proved exceptionally powerful.  However, until recently it has only been studied
	in two dimensions --- either through boost-invariant simulations or studying only quantities at mid-rapidity.   We explore an extension to 3 dimensions, in order to determine whether a similar understanding can be obtained for the rapidity dependence of the collision system.  In particular, we introduce rapidity-dependent eccentricities and investigate a trivial extension of the 2D eccentricity scaling of elliptic and triangular flow, as well as a way to systematically improve these initial-state estimators.  We then explore the dependence of the resulting new response coefficients on shear viscosity and initial total energy.
	\end{abstract}

\maketitle
	
	\section{Introduction}

		The evolution of a relativistic heavy-ion collision is a complicated process, well described by simulations that utilize relativistic fluid dynamics. However, despite the non-linear nature of these equations, simple relations can be found between the state of the system at the onset of hydrodynamic behavior, and the final spectrum of detected particles \cite{Gardim:2011xv}.  In the simplest case a linear relation between a particular measure of the initial spatial azimuthal asymmetry $\varepsilon_n$ and the final momentum anisotropy $V_n$  can be written as the vector relation
		\begin{equation}
		\label{lin}
		V_n = \kappa_n \varepsilon_n,
		\end{equation}
		where all relevant information about the initial state is contained in a single vector quantity $\varepsilon_n$, and all information about the subsequent dynamics of the system is contained in the response coefficient $\kappa_n$.  This separation of effects is quite powerful, and has allowed for numerous insights \cite{Kolb:2001qz, Bhalerao:2011yg, Luzum:2012wu, Retinskaya:2013gca, Gardim:2018lao}. 
		
		However, these relations were developed in systems with two effective dimensions --- first in systems with an assumed invariance under longitudinal boosts \cite{Kolb:2001qz}, and later in 3D simulations, but analysis restricted to mid-rapidity \cite{Gardim:2011xv}.  Only recently has an extension to a full 3 dimensions has been considered \cite{Li:2019eni}.
		
		Here we investigate possible extensions of this mapping of the initial to final state, to include the dependence on rapidity.
		
	\section{Initial to final state mapping}
	    The essential idea of the mapping from initial to final state is the ansatz that hydrodynamic evolution of the system is more sensitive to the large-scale structure of the initial conditions, as compared to features at small length scales.  
	    
	    This separation of scales is accomplished via Fourier transform of the initial energy density $\rho$, and large length scales isolated via a truncated Taylor series around transformed variable $k = 0$.  That is, \cite{Teaney:2010vd}
	    
		\begin{align}
		\rho(\vec{k}_{\perp}) &= \int d^2x_{\perp} \rho(\vec{x_{\perp}}) e^{i \vec{k_{\perp}}\cdot \vec{x_{\perp}}}, \\
		W({\vec k_{\perp}}) \equiv 
		\ln\left[\rho(\vec{k_{\perp}})\right] &=  \sum_{n=-\infty}^\infty  \sum_{m=|n|}^\infty W_{n,m} k_{\perp}^m e^{-i n\phi_k } .
		\end{align}
		
		The truncated set of complex cumulants $\{W_{n,m}\}$ are then used to construct initial state estimators of the final distribution of particles in momentum space, which are traditionally separated into azimuthal rotation modes
		\begin{align}
		\label{dist}
		\frac{dN}{d\phi} &= {{N}\over{2\pi}}  \sum_{n=-\infty}^\infty V_n e^{-in\phi},
		\end{align}
		where $\phi$ is the azimuthal angle of a particle's momentum.
		
		The leading order relation is a linear proportionality between $V_n$ and the lowest cumulant $W_{n,n}$.  Typically, one constructs dimensionless ``eccentricities''; for example,
		\begin{equation}
		\varepsilon_n = -n!\frac{W_{n,n}}{W_{0,2}^{n/2}}.
		\end{equation}
		In  a coordinate system such that $W_{1,1}=0$, this has the explicit form for $n=2,3$
		\begin{equation}
		\label{ecc}
		\varepsilon_n = - \frac{\langle r_\perp^n e^{in\varphi}\rangle}{\langle r_\perp^2\rangle^{n/2}},
		\end{equation}
		with
		\begin{equation}
		\langle \ldots \rangle \equiv \frac{\int d^2x_{\perp}  \rho(\vec x_{\perp})\ldots}{\int d^2x_{\perp}  \rho(\vec x_{\perp})}.
		\end{equation}
		
		The resulting linear relation \eqref{lin} between $V_n$ and $\varepsilon_n$ has been shown to be an excellent approximation for $n = 2$ and $n=3$ \cite{Gardim:2011xv}.
		Other harmonics can have significant contributions from non-linear combinations of cumulants, so for simplicity we focus here on these two harmonics.  The extension to estimators with non-linear or higher order cumulant terms is straightforward.
		
	\section{Extending to 3D}
		
		All the above relations are implicitly 2-dimensional, representing dynamics in the transverse plane, and with the longitudinal dimension ignored.  The justification for this is that, at the highest collision energies, the system should have an approximate invariance  under longitudinal boosts.  That is, the initial energy density $\rho$ should be approximately independent of spacetime rapidity
		\begin{align}
		\eta \equiv \arctan \frac{z}{t}
		\end{align}
		and the final particle distribution \eqref{dist} independent of rapidity
		\begin{equation}
		Y \equiv \arctan \frac{p_z}{E}
		\end{equation}
		such that
		\begin{align}
		\rho(\vec x_\perp, \eta)&\simeq \rho(\vec x_\perp),\\
		\frac{dN}{d^3p}(\vec p_\perp, Y) &\simeq \frac{dN}{d^3p}(\vec p_\perp).
		\end{align}
		in fact, boost invariant simulations have been quite successful at describing experimental data near mid-rapidity, $Y\simeq 0$ \cite{Ryu:2017qzn, Niemi:2015qia}.
		
		Nevertheless, numerous rapidity-dependent measurements have been made, which can be utilized to gain insight into QCD dynamics.   Because of this, it is of great interest to extend the 2D relations to 3D, to give a better understanding of the evolution of the system and to provide guidance for how to separate, e.g., initial state and final state effects.
		
		To that end, we must consider the dependence of $V_n$ on rapidity, and construct estimators from the initial state that contain information about spacetime rapidity $\eta$.
		
		A straightforward way to do this is to simply calculate eccentricities independently at each value of rapidity $\eta$, to construct rapidity-dependent eccentricities:
		\begin{align}
		\varepsilon_n(\eta) &= - \frac{\langle r_\perp^n e^{in\varphi}\rangle_\eta}{\left(\langle r_\perp^2\rangle_\eta\right)^{n/2}},\\
		\langle \ldots \rangle_\eta &\equiv \frac{\int d^2x_{\perp}  \rho(\vec x_{\perp},\eta)\ldots}{\int d^2x_{\perp}  \rho(\vec x_{\perp},\eta)}.
		\end{align}
		
		To first approximation, we might expect the results to be local in rapidity.  That is
		\begin{align}
		V_n(Y) = \kappa_n \varepsilon_n(\eta = Y).
		\end{align}
		
		Of course, we don't expect the results to be perfectly local in rapidity.  Hydro evolution can propagate information from one rapidity $\eta$ in the initial state to another rapidity at later times.  In addition, particles in a fluid (e.g., at freeze-out) do not all have rapidity $Y$ equal to the spacetime rapidity $\eta$ of the fluid cell, but instead have a distribution of finite extent.  Similarly, unstable particles can decay to daughter particles that do not have exactly the same rapidity as the parent.  
		
		Nevertheless, we do expect a quasi-locality.  All of these effects should be largest at small rapidity difference $|\eta - Y|$, with an exponential decrease such that properties of the initial state at far forward rapidity $\eta$ have essentially no effect on the particle distribution at far backward rapidity $Y$.
		
		This decreasing effect can be encoded in a gradient expansion:
		\begin{align}
		\label{grad}
		V_n(Y) = \kappa_n \varepsilon_n(Y) + 
		\kappa''_n \varepsilon_n''(Y) + \kappa_n'''' \varepsilon_n''''(Y) + \ldots,
		\end{align}
		where the primes on the eccentricities indicate a derivative with respect to spacetime rapidity $\eta$ and where the eccentricity and its derivatives are then evaluated at $\eta=Y$.
		Note that symmetry dictates that the hydro response functions must be symmetric under reflection, and so only even derivatives of $\eta$ can appear. 
		
	\section{Numerical tests}
		
		In order to determine whether these estimators describe the evolution of a heavy-ion collision system, we perform  hydrodynamic simulations.  For this we use the MUSIC \cite{Schenke:2010nt, Schenke:2010rr, Paquet:2015lta} 3+1D hydrodynamic code.    Specifically, all simulations use the equation of state s95p-v1 and vanishing bulk viscosity. The initial viscous tensor is set to 0, the initial fluid velocity is set to boost-invariant Bjorken flow.  Freeze-out occurs at an energy density of 0.12 $GeV\over{fm^3}$ (equivalent to temperature of 137 $MeV$).   
		
		We aim to study the hydrodynamic response to initial conditions.  To that end, we construct a parameterized toy initial condition that can be varied by hand in order to probe the resulting response.  Specifically, we choose a simple Gaussian energy density distribution at an initial time $\tau = 1 fm$, which can be deformed in the desired rotational harmonic $n$:
		\begin{equation}
			\rho(\vec x_\perp, \eta) = A(\eta) \exp\left(\frac{-r^2}{2\rho^2}\left[1+\bar\varepsilon_n(\eta) \cos(n\varphi)\right]\right)
			\label{eq:ED2}
		\end{equation}
		
		The parameter $\bar \varepsilon_n$ controls the azimuthal asymmetry, such that the eccentricity can be directly controlled
		\begin{align}
		\varepsilon_2 = \bar\varepsilon_2,
		\end{align}
		while the triangularity can also be easily controlled,  having the approximate relation
		\begin{align}
		\varepsilon_3 \simeq 1.65 \bar\varepsilon_3.
		\end{align}
		In general, we will allow the normalization $A$ and asymmetry $\varepsilon_n$ to vary with rapidity $\eta$.  Flow observables are calculated for direct pions, neglecting the effect of resonance decays.

	\section{Results and discussions}
		First we perform ideal (zero viscosity) boost-invariant simulations, with all parameters independent of rapidity $\eta$.  This allows us to verify the traditional linear relationship \eqref{lin}, but also to extract the response coefficients $\kappa_n$, which should remain the same when the system is boost variant. 
		
		In the following, we fix $\rho = 3 fm$ and $A = 50 fm^{-4}$, corresponding roughly to a typical collision at RHIC.  By varying the asymmetry parameters $\varepsilon_n$, we indeed find a linear behavior, from which we can extract values
		\begin{align}
			\kappa_2 &= 0.174\\
			\kappa_3 &= 0.068.
		\end{align}
		
\begin{figure}
			\includegraphics[width=\linewidth]{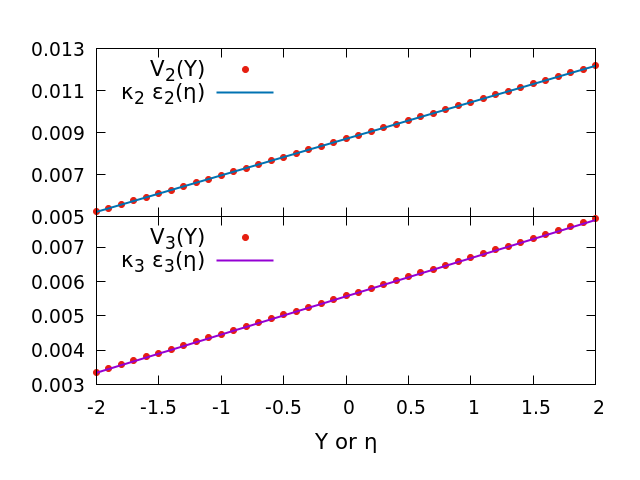}
			\caption{\label{fig:VnEnLinear} Elliptic flow $V_2(Y)$ and triangular flow $V_3(Y)$ as a function of rapidity $Y$, plotted against the corresponding linear estimator $\kappa_n \varepsilon_n(\eta)$ for the case of a linear dependence on  $\eta$,  with $\kappa_n$ taken from corresponding boost-invariant simulations:  $\kappa_2 = 0.174$, $\kappa_3 = 0.068$.   
			}
		\end{figure}

		With these values in mind, we can finally explore rapidity-dependent systems.  For example, we can choose a linear function $\varepsilon_n(\eta)$, and compare it to the resulting rapidity-dependent $V_n(Y)$, as shown in Fig.~\ref{fig:VnEnLinear}.  Using the same $\kappa_n$ as found in the boost invariant case, we find that the estimator $\eqref{lin}$ provides an excellent description of the results.  Not only is the result a linear function of rapidity, but the proportionality constant between $V_n$ and $\varepsilon_n$ is the same as found previously.  We found this to be true independent of the slope of the rapidity dependence.

		Note that, in this case, all higher terms in Eq.~\eqref{grad} vanish, since all even derivatives of $\varepsilon_n(\eta)$ vanish.  To test to what extent this gradient expansion describes the hydrodynamic response, we can include curvature to the initial eccentricity.
		
\begin{figure}
			\includegraphics[width=\linewidth]{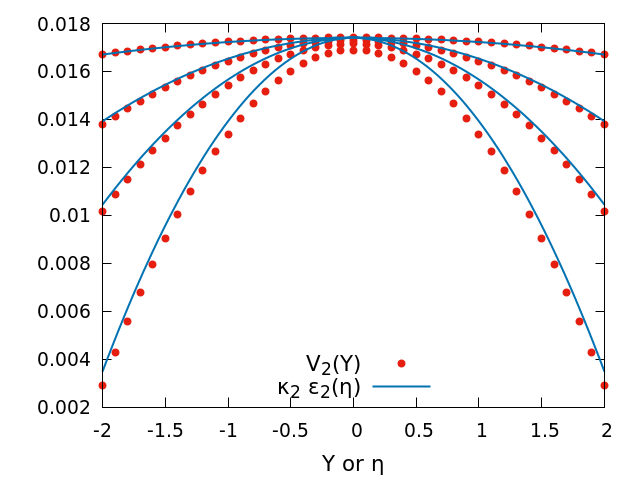}
			\caption{\label{fig:QuadraticPlot} 
			Anisotropic flow $V_2(Y)$ compared to the first-order estimator $\kappa_2 \varepsilon_2(\eta)$ with a quadratic dependence on $\eta$, and using the response coefficient from the boost-invariant analysis, $\kappa_2 = 0.174$.
			}
		\end{figure}

		In Fig.~\ref{fig:QuadraticPlot}, we show the corresponding result for $\varepsilon_2(\eta)$ with quadratic dependence on rapidity and various values for the curvature, compared to the first-order estimator \eqref{lin}.

%
%
%
%
		
		We observed the same behavior for $n=3$.
		The first-order estimator $\eqref{lin}$ continues to give a reasonable description of the results, but the accuracy gets poorer as the curvature increases.  In fact, a close inspection shows that the difference between the anisotropic flow and the estimator is constant in rapidity, but grows proportionally to the curvature of the initial eccentricity.  This is exactly what is expected from the proposed gradient expansion \eqref{grad}, which in this case has exactly 2 terms:
		\begin{equation}
		\label{second}
		V_n(Y) = \kappa_n\varepsilon_n(\eta=Y) + \kappa_n''~\varepsilon_n''(\eta=Y).
		\end{equation}
		From these results we can extract the values of $\kappa_n''$.  We find
		\begin{align}
		\kappa_2'' = 0.0135,\\
		\kappa_3'' = 0.0092.
		\end{align}
		The gradient corrections are small, as expected, but not necessarily negligible, depending on the curvature of the initial eccentricity.  We observe the same behavior for $n=3$ (not shown).
%
%

\begin{figure}
			\includegraphics[width=\linewidth]{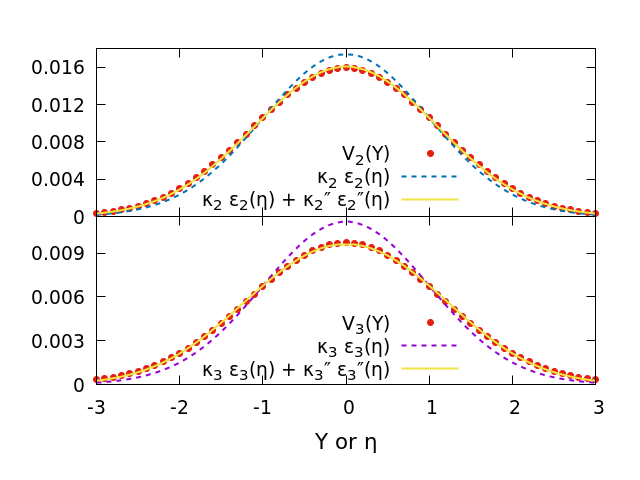}
			\centering
			\caption{\label{fig:GaussianPlot} 
			Anisotropic flow $V_n(Y)$ compared to the first-order and second-order estimators with a Gaussian dependence on $\eta$, and using response coefficients from previous analyses, $\kappa_2 = 0.174$, $\kappa_3 = 0.068$, $\kappa_2'' = 0.0135$, $\kappa_3'' = 0.0092$.
			}
		\end{figure}

		Next, in Fig.~\ref{fig:GaussianPlot} we consider a Gaussian rapidity dependence.  In this case, an infinite number of derivatives are present, and we can probe Eq~\eqref{grad} in more detail.  Again, we see that the leading-order estimator \eqref{lin} offers a reasonable description of the resulting anisotropic flow, but a better description can be obtained by adding a subleading term, Eq.~\eqref{second}, with the same response coefficient $\kappa''_n$ as found in the quadratic simulation.   We find that the contribution from the next correction (from the fourth derivative) is suppressed even more than the second order term, such that it is negligible in this case.

		
		\subsection{Viscosity dependence}
\begin{figure}[h]
			\includegraphics[scale=0.4]{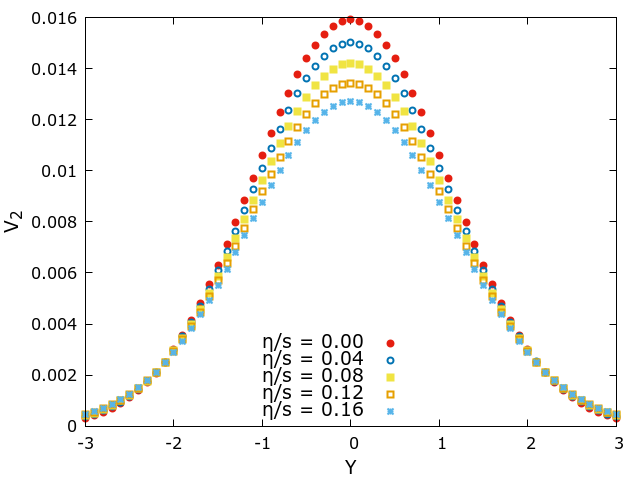}
			\centering
			\caption{\label{fig:GaussianViscosityPlot} Compilation of graphics of $V_2$ in function of rapidity $Y$ for different values of shear viscosity $\frac{\eta}{s}$}
		\end{figure}

		Next, we include a non-zero shear viscosity, in order to probe the behavior of the new response coefficients $\kappa_n''$.  Here, we use a Gaussian rapidity dependence, exactly as in Fig.~\ref{fig:GaussianPlot}, in which case we can extract both $\kappa_n$ and $\kappa_n''$ from a single simulation.  Figure \ref{fig:GaussianViscosityPlot} shows the resulting rapidity-dependent elliptic flow for various values of shear viscosity (with $\eta/s$ constant).

		
\begin{figure}
			\includegraphics[width=\linewidth]{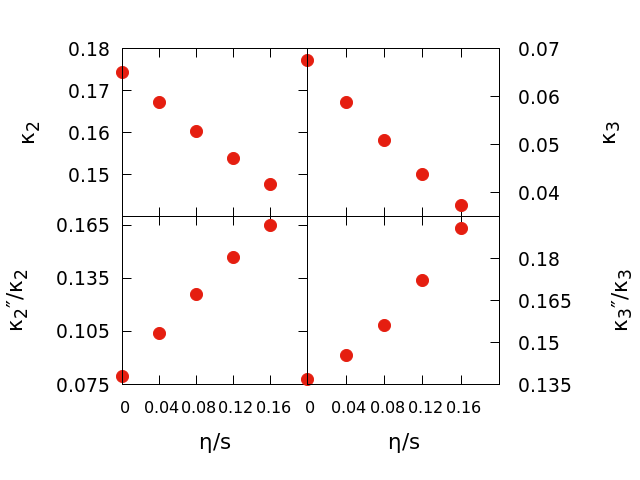}
			\caption{\label{fig:KK''xViscPlot} Response coefficients $\kappa_n$ and the ratio ${\kappa_n''}/{\kappa_n}$ as a function of shear viscosity $\frac{\eta}{s}$}
		\end{figure}

		As expected, viscosity suppresses anisotropic flow, represented by a decrease in leading response coefficient $\kappa_n$.  This dependence is shown explicitly in the top panels of Fig.~\ref{fig:KK''xViscPlot}.

%
		
		The bottom panels of Fig.~\ref{fig:KK''xViscPlot} show an interesting new result.  The subleading response coefficients' ratios ${\kappa_n''}/{\kappa_n}$ increase with viscosity, indicating an increased sensitivity to rapidity regions $\eta\neq Y$ when viscosity is larger.
		
		

		We have thus verified the ansatz \eqref{grad}, and investigated the dependence of new response coefficients to viscosity.  However, in realistic simulations, not only the asymmetry depends on rapidity, but also the total energy and multiplicity.   An important question is whether in this case we can still understand the hydrodynamic response to initial conditions in a simple way.
		
		\subsection{Rapidity dependence via energy dependence}
		Even in boost-invariant simulations, the anisotropic flow $V_n$ resulting from an initial condition with spatial anisotropy $\varepsilon_n$ will change if the total energy/entropy of the initial condition changes significantly.  That is, the hydrodynamic response $\kappa_n$ depends on initial energy.  The natural question is whether this effect alone can describe the hydrodynamic response of systems with rapidity-dependent initial energy.
		
\begin{figure}
			\includegraphics[width=\linewidth]{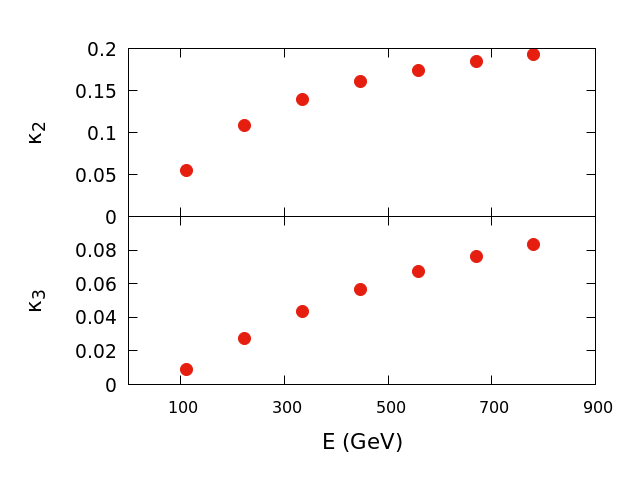}
			\caption{\label{fig:KK''xEPlot} Response coefficient $\kappa_n$ as a function of total energy $E$ in $GeV$}
		\end{figure}

		To verify this, we first calculate $\kappa_n(E)$ in boost-invariant simulations, where in our initial condition the total energy per unit rapidity is given by
		\begin{equation}
			E(\eta) = \tau\int \rho(\vec x_\perp, \eta) d^2x_\perp = \frac{2\pi A\rho^2\tau}{\sqrt{(1-\bar\varepsilon_{n}^2)}}.
			\label{eq:TotalE}
		\end{equation}
		with $\tau = 1 fm$ the initial proper time. 
The result can be seen in Fig.~~\ref{fig:KK''xEPlot}.
		
		Then we verify whether the rapidity-dependent results are described by
		\begin{align}
		\label{energy}
		V_n(Y) \simeq \kappa_n\left[E(\eta=Y)\right] \varepsilon_n.
		\end{align}
		
\begin{figure}
			\includegraphics[width=\linewidth]{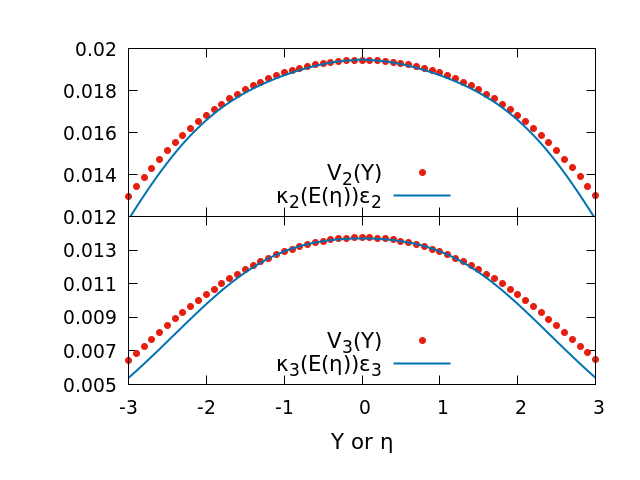}
			\caption{\label{fig:EnergyModelBIPlot} Harmonic flows $V_n(Y)$ compared to the estimator with response coefficient $\kappa_n$ as a function of total energy $E$ which is dependent on spatial-rapidity times boost-invariant eccentricities: $\varepsilon_2 = 0.1$ and $\varepsilon_3 = 0.165$}
		\end{figure}

		We choose a Gaussian profile for the initial energy as a function of rapidity and compute the resulting elliptic and triangular flows. In  Fig.~\ref{fig:EnergyModelBIPlot} we can see that the behavior is reasonably captured by Eq.~\eqref{energy}, and locality in rapidity can indeed be maintained.
		
It should be noted, however, that in the case where both the initial energy and initial eccentricity depend on rapidity, the symmetry argument that guided the form of Eq.~\eqref{grad} no longer holds.   That is, one can now form reflection-symmetric terms involving first derivatives.  We explore this extension in Appendix~\ref{reflect}.

		\section{Conclusions}
		
		We showed that the traditional eccentricity scaling of heavy-ion collisions can be extended to a rapidity-dependent form, and systematically improved to include non-local influence in rapidity.  We used hydrodynamic simulations to verify this framework, and to extract the viscosity and total energy dependence of newly-introduced response coefficients.
		
		We found that an increase in viscosity makes elliptic flow and triangular flow less dependent on the local value of eccentricity (i.e., the eccentricity at spacetime rapidity $\eta$ equal to particle rapidity $Y$ of the flow in question), and more sensitive to the eccentricity in the neighborhood surrounding rapidity $Y$.  
		
		A complementary ansatz is presented in Appendix~\ref{app}.
		
		In realistic systems, the total energy of the system will have a non-trivial dependence on rapidity as well.  We find that locality in rapidity is maintained as long as the dependence of the response coefficients $\kappa_n$ on energy is taken into account.
		
		In Appendix~\ref{reflect} we discuss a new term on the gradient expansion of the eccentricities \eqref{grad} involving a product of first derivatives of initial energy and of initial eccentricity.
		
		These results should allow for new insight into the dynamics of heavy ion collisions, such as a more direct isolation of rapidity-dependent processes in the initial stages.

\section*{Acknowledgments}

This work was supported by FAPESP projects 2016/24029-6, 2017/05685-2 and 2018/24720-6, and by project INCT-FNA Proc.~No.~464898/2014-5.  R.S.~acknowledges support from FAPESP projects 2017/18632-4 and  2019/04318-1.
		
		\appendix
		
		\section{An alternative ansatz}
		\label{app}
		
		Rather than assuming an expansion in gradients, $\eqref{grad}$, one might try the following ansatz \cite{Li:2019eni}:
		\begin{align}
		\label{eq:ModelWeight}
			V_n(Y)  &= \int_{-\infty}^{\infty} d\eta\, W_n(Y-\eta) \varepsilon_n(\eta), 
		\end{align}
		where the effects of eccentricity $\varepsilon_n$ at each space-time rapidity $\eta$ influences the anisotropic flow $V_n$ at rapidity $Y$ additively, according to the kernel $W_n(Y-\eta)$.
		A perfectly local relation, Eq.~\eqref{lin}, corresponds to $W_n(Y-\eta) \propto \delta(Y-\eta)$, though in general one expects a function with finite width.
		
		In principle, one could find the function $W_n$ by setting $\varepsilon_n(\eta) = \delta(\eta)$, in which case
		\begin{align}
		V_n(Y) &= \int_{-\infty}^\infty d\eta W_n(Y-\eta) \delta(\eta) = W_n(Y)
		\end{align}
		
		In practice, we can approach this limit by choosing a function with small width.  Here we try a step function for $\varepsilon(\eta)$ with a width the size of one hydro cell, in our case $\Delta\eta = 0.1$.
		
\begin{figure}
			\includegraphics[scale=0.4]{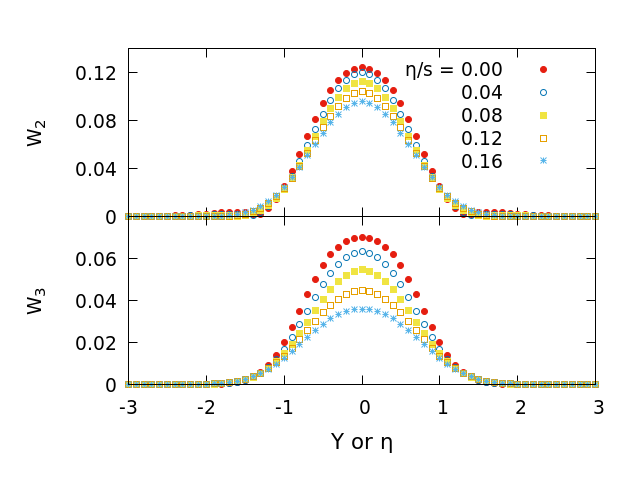}
			\caption{\label{fig:DeltaPlot} Resulting values of $W_n$ for each value of shear viscosity $\eta/s$ in function of rapidity $Y$ for $\alpha=0.1$ and $\Delta\eta=0.1$}
		\end{figure}

		Then, noting the discrete grid in $\eta$, we have
		\begin{align}
		V_n(Y) &= \sum_{\eta} W_n(Y-\eta)  \alpha\delta_{\eta,0} \Delta\eta.\\
		&= \alpha W_n(Y) \Delta\eta
		\end{align}
		We show the resulting $W_2$ and $W_3$ in Fig.~\ref{fig:DeltaPlot}, for various values of shear viscosity.
		
		We can see that the influence of $\varepsilon_n(\eta)$ extends not much more than 1 unit of rapidity, with a smooth shape (that is not in general well described by a Gaussian).
		
		If the ansatz is correct, we should be able to make contact with the previous results from Eq.~\eqref{grad} by expanding $\varepsilon_n(\eta)$ in a Taylor series around $Y= \eta$,
%
%
%
%
%
%
		\begin{equation}
			\varepsilon_{n}(\eta) = \varepsilon_{n}(\eta)\bigg|_{\eta=Y} + \frac{1}{2} \frac{\partial^2 \varepsilon_{n}}{\partial \eta^2}\bigg|_{\eta=Y}(\eta-Y)^2 + ...
			\label{eq:EnExpansion}
		\end{equation} 
		and substituting Eq.~(\ref{eq:EnExpansion}) in Eq.~(\ref{eq:ModelWeight}). 
		We find

		\begin{align}
		\kappa_n &= \int d\eta \,W_n(\eta)\\
		\kappa''_n &= {1 \over 2} \int d\eta\, (\eta )^2W_n(\eta)
		\end{align}
		
		where $\kappa_n$ and $\kappa''_n$ are the previous response coefficients.
		
\begin{figure}
			\includegraphics[scale=0.4]{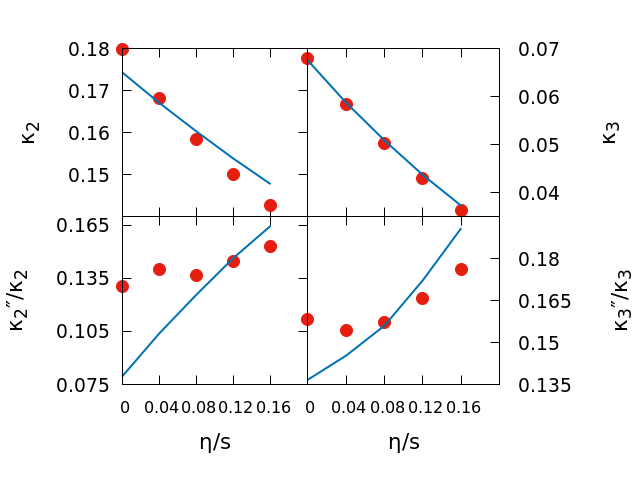}
			\caption{\label{fig:KK''WeightPlot} Resulting values of $\kappa_n$ and the ratio ${\kappa_n''}/{\kappa_n}$ for each value of shear viscosity $\eta/s$ and the previously obtained ones represented in lines for comparison}
		\end{figure}

		Using the results of Fig.~\ref{fig:DeltaPlot}, we compute these response coefficients as a function of viscosity.  The results are shown in Fig.~\ref{fig:KK''WeightPlot}.  The coefficients $\kappa_n$ behave as expected, and agree reasonably well with the previous results of Fig.\ref{fig:KK''xViscPlot}.  The ratios ${\kappa''_n}/{\kappa_n}$ are not equal, but qualitatively similar.
		
		Thus, the ansatz \eqref{eq:ModelWeight} appears to be in agreement with previous results, and could give additional insight into the hydrodynamic response.  Better precision could likely be obtained by significantly reducing the cell size so that the initial condition can be just as narrow, yet smooth instead of a discontinuous step.
		
		\section{Contribution from first derivatives}
		\label{reflect}
		
		The reflection symmetry of the hydrodynamic response demands that each term in the estimator is even under reflection $\eta\to -\eta$.   First derivatives with respect to space-time rapidity are odd under reflection, but a product of two such quantities maintains the correct symmetry.  Specifically one can have a contribution from the derivative of initial energy and that of initial eccentricity.  Our estimator  can thus have the form
%
%
		\begin{equation}
		\label{eq:newterm}
		V_n(Y) = \kappa_n \varepsilon_n (Y) + \kappa'_n {E'(Y)\over{E(Y)}} \varepsilon'_n(Y) + \kappa''_n \varepsilon''_n(Y).
		\end{equation}
		
We can isolate the effect of the new term by performing simulations with linear dependences on rapidity.    We verify that, when varying the slopes of initial energy and eccentricity, we obtain consistent results and a value for the response coefficient of
%
%
		\begin{eqnarray}
			\kappa'_2 = 0.063.
			\\
			\kappa'_3 = 0.035.
		\end{eqnarray}
%
%

Thus, we have $\kappa'_n/\kappa''_n \simeq$ 4--5, so that the first-derivative term is comparable to the previous second-derivative term only when
\begin{align}
\label{eq:k'2/k''2}
5\left|\frac{E'}{E}\right| \sim \left|
 \frac{\varepsilon''_n}{\varepsilon'_n}\right|.
\end{align}

\begin{figure}
	\includegraphics[scale=0.4]{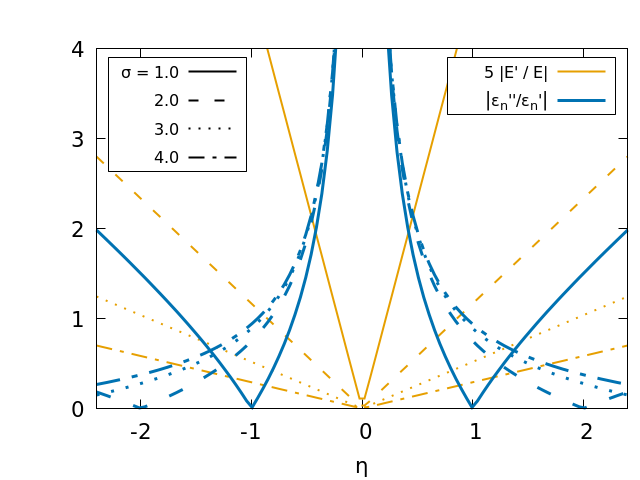}
	\caption{\label{fig:LHS_RHS} Comparison between values of the left-hand side (yellow) and right-hand side (blue)
		 of Eq.~\eqref{eq:k'2/k''2}, assuming Gaussian dependence on rapidity with width 1 (solid lines),  2 (dashed),  3 (dotted), and 4 (dash-dotted).
	 }
\end{figure}

To illustrate this relation, in Fig.~\ref{fig:LHS_RHS}
%
 we plot the left- and right-hand side for various distributions of initial energy and eccentricity.  Specifically, we choose a Gaussian profile
 \begin{align}
 f(\eta) &= e^{-\frac{\eta^2}{2\sigma^2}},
 \end{align}
 so that if we choose a Gaussian eccentricity profile $\varepsilon_n(\eta) \propto f(\eta)$, the right-hand side of Eq.~\eqref{eq:k'2/k''2} becomes
   \begin{align}
 \left|\frac{\varepsilon''_n}{\varepsilon'_n}\right|&= \left|\frac{1}{\eta} - \frac{\eta}{\sigma^2}\right|
 \end{align}
 and if the energy profile is Gaussian $E(\eta)\propto f(\eta)$ the left-hand side of Eq.~\eqref{eq:k'2/k''2} becomes
 \begin{align}
5 \left| \frac{E'}{E}\right|&= 5\frac{|\eta|}{\sigma^2}.
 \end{align}
 

%

%

We  find that typically the left side is much smaller, and therefore the second term of  Eq.~\eqref{eq:newterm} is subdominant, except at an inflection point of $\varepsilon_n(\eta)$. However, this is dependent on the true initial conditions, which are not known.  Additionally, the relative value of response coefficients $\kappa'/\kappa''$ could change somewhat with a more realistic treatment or with different values of  hydrodynamic parameters.

\end{document}